\documentclass[conference]{IEEEtran}
\IEEEoverridecommandlockouts

\usepackage{cite}
\usepackage{amsmath,amssymb,amsfonts}
\usepackage{algorithmic}
\usepackage{graphicx}
\usepackage{textcomp}
\usepackage{multirow}
\usepackage{multicol}
\usepackage{booktabs}
\usepackage{amsmath}
\usepackage{amssymb}
\usepackage{xcolor}
\usepackage{subfiles}
\usepackage{balance}
\usepackage[colorlinks=True]{hyperref}

\def\BibTeX{{\rm B\kern-.05em{\sc i\kern-.025em b}\kern-.08em
    T\kern-.1667em\lower.7ex\hbox{E}\kern-.125emX}}

\newcommand{\Href}[1]{
	\href{#1}{\scriptsize #1}
}

\begin{document}

\title{Generalized Audio Deepfake Detection Using Frame-level Latent Information Entropy\\
	\author {
		Botao Zhao$^{*}$,
		Zuheng Kang$^{*}$,
		Yayun He$^{*}$,\thanks{* These authors contributed equally to this work.}
		Xiaoyang Qu,
		Junqing Peng,
		Jing Xiao,
		Jianzong Wang$^{\dagger}$\thanks{$^{\dagger}$ Corresponding author, jzwang@188.com,} \\
		Ping An Technology (Shenzhen) Co., Ltd., Shenzhen, China. \\
	}

}

\maketitle
\vspace{-0.7cm}
\begin{abstract}
	Generalizability, the capacity of a robust model to perform effectively on unseen data, is crucial for audio deepfake detection due to the rapid evolution of text-to-speech (TTS) and voice conversion (VC) technologies.
	A promising approach to differentiate between bonafide and spoof samples lies in identifying intrinsic disparities to enhance model generalizability.
	From an information-theoretic perspective, we hypothesize the information content is one of the intrinsic differences: bonafide sample represents a dense, information-rich sampling of the real world, whereas spoof sample is typically derived from lower-dimensional, less informative representations.
	To implement this, we introduce frame-level latent information entropy detector(f-InfoED), a framework that extracts distinctive information entropy from latent representations at the frame level to identify audio deepfakes.
	Furthermore, we present AdaLAM, which extends large pre-trained audio models with trainable adapters for enhanced feature extraction.
	To facilitate comprehensive evaluation, the audio deepfake forensics 2024 (ADFF 2024) dataset was built by the latest TTS and VC methods.
	Extensive experiments demonstrate that our proposed approach achieves state-of-the-art performance and exhibits remarkable generalization capabilities.
	Further analytical studies confirms the efficacy of AdaLAM in extracting discriminative audio features and f-InfoED in leveraging latent entropy information for more generalized deepfake detection.
\end{abstract}


\section{Introduction}
\label{sec:intro}

Distinction between synthesized and real speech is becoming increasingly narrow, Because of the rapid development of text-to-speech (TTS) \cite{kim2021conditional,zhao2022nnspeech} and voice conversion (VC) technologies\cite{10842513, deng2023pmvc}.
Thanks to many highly sophisticated TTS and VC projects that have been open-sourced, these highly realistic synthesis techniques are very easily available to everyone.
However, the undesirable development and malicious spread of these technologies have also enhanced the potential for misuse, thereby posing a significant threat to social stability and public security.
Therefore, detection of deepfake audio has become a critical imperative for avoiding technology misuse.

The relationship between deepfake synthesis and deepfake detection can be likened to that of two siblings, each evolving as the other progresses.
In the realm of audio deepfake detection, recent years have witnessed the emergence of diverse methodologies \cite{jung2022aasist,guo2024audio, kang2024retrieval, zhang2024remember, wang2024multi}.
However, existing detection models typically exhibit suboptimal performance when confronted with novel datasets or audio generated by previously unseen techniques, a phenomenon termed ``lack of generalization capability'' \cite{muller2022does, yi2023audio}.
To address this, researchers have pursued various strategies.
One approach involves enhancing generalizability by exploiting the pre-trained large audio models \cite{yang2024robust, guo2024audio, lu2024one}.
Another focuses on domain adaptation \cite{xie2023learning}.
Additionally, the continual learning approach enables detection models to incrementally absorb knowledge from newly acquired data \cite{zhang2024remember}.
Despite these advancements, generalized audio deepfake detection has not yet to be achieved.
\textit{Therefore, this study endeavors to uncover a fundamental distinction between bonafide and spoof samples, facilitating the development of a generalized detection method.}
To enhance the generality, it is essential to identify common features that can distinguish bonafide sample from spoof audio.
Bonafide audio originates from the real world, retaining rich information despite minor losses during air transmission and minimal reduction through microphone recording and digital sampling.
In contrast, TTS/VC methods, typically based on encoder-decoder structures, generate spoofed audio from highly compressed latent vectors with less information. While these methods aim to reconstruct audio via mel-spectrograms or raw waveforms, their optimization focuses on human hearing, often discarding imperceptible high-frequency signals.
\textit{Therefore, we hypothesize that the main difference between bonafide and spoof audios lies in the amount of information they hold, a characteristic that is prevalent in a variety of generation methods.}

Information entropy, a classical concept proposed by Shannon \cite{shannon1948mathematical}, has proven to be an efficient approach for quantifying the information content of a system in information theory and exerts a profound influence on the current information era.
The concept of entropy is widely adopted as a loss function in deep learning, such as cross-entropy and mutual information.
Recently, information-theoretic approaches linking generalization capabilities to compression have gained increasing attention \cite{gabrie2018entropy}.
However, most studies focus primarily on theoretical research \cite{chechik2003information}. \cite{wang2024multi} proposed a multi-scale signal statistic utterance-level entropy permutation method for audio deepfake detection, which lead to coarse-grained classification.
Unlike the previous works, our study attempts to combine frame-level information entropy theory in addressing the deepfake detection problem.
\begin{figure}[t]
	\centering
	\includegraphics[width=0.85\linewidth]{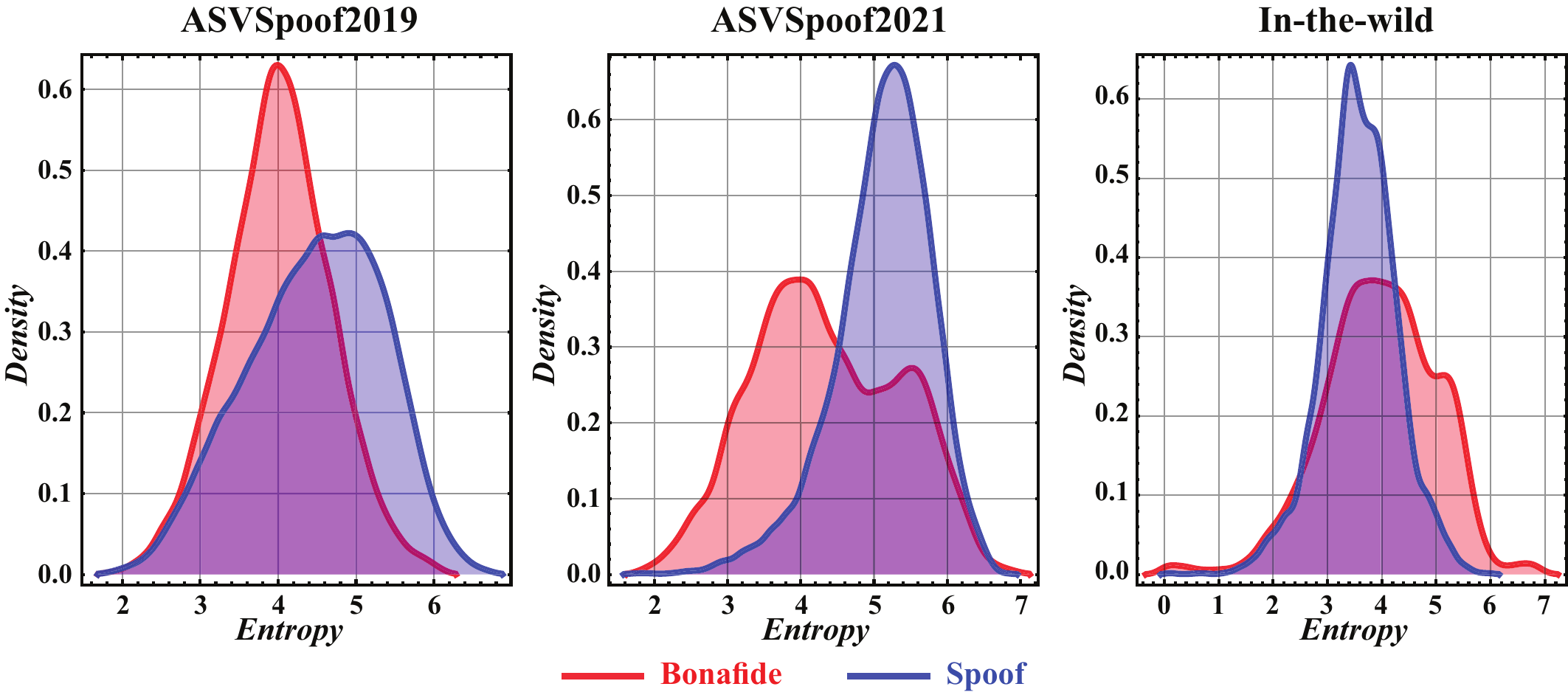}
	\vspace{-0.9em}
	\caption{The distribution of entropy on ASVspoof 2019 LA, ASVspoof 2021 DF, and In-the-wild dataset.}
	\label{fig:entropy_distribution}
	\vspace{-1.6em}
\end{figure}

To introduce the information entropy theory into the audio deepfake detection, however, there are several challenges:
Firstly, information entropy is sensitive to noise, which is inevitable in bonafide audio.
Secondly, audio contains diverse content and the different segments could have various information entropy, which will amplify the difference within the same class.
Fig. \ref{fig:entropy_distribution} provides an initial glimpse into the information entropy across various datasets.
A statistical analysis of the entropy distribution of the audio training sets from ASVspoof datasets revealed a notable difference between the bonafide and spoof audios, but this distinction was less pronounced in the In-the-wild dataset.
Therefore, to implement this idea, we have made the following contributions:
\begin{itemize}
	\item We proposed the frame-level latent information entropy detector (f-InfoED), a novel approach that achieves generalized audio deepfake detection from an information-theoretic perspective.
	\item We released the audio deepfake forensics 2024 (ADFF 2024) dataset, which was generated by the most recent TTS/VC methods for evaluation.
	\item Experiments shows that our proposed method not only achieves SOTA performance in cross-dataset audio deepfake detection, but also has the generalization capability to unseen perturbations and other modalities.
\end{itemize}

\begin{figure*}[t]
	\centering
	\includegraphics[width=0.9\linewidth]{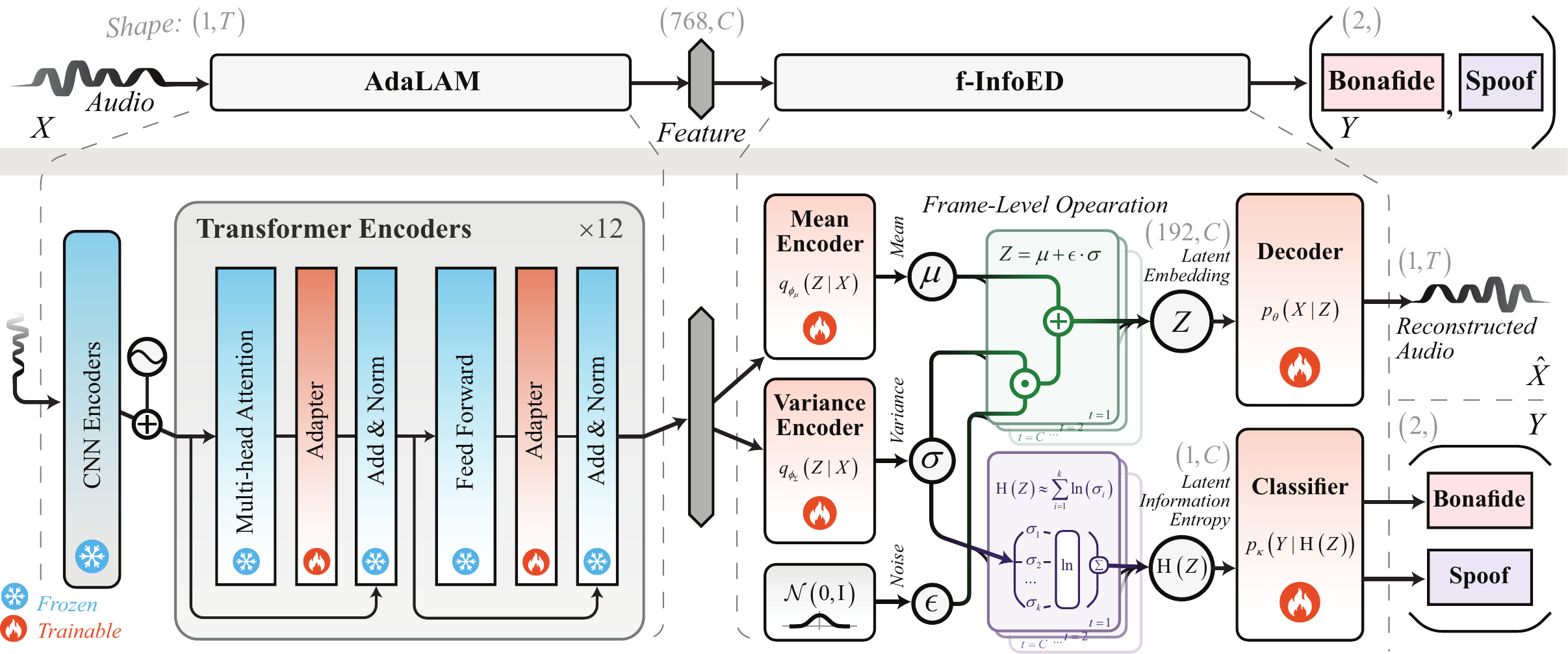}
	\caption{The overall pipeline of our proposed generalized audio deepfake detection method.}
	\label{fig:overall}
\end{figure*}

\section{Methodology}
Our proposed framework (Fig. \ref{fig:overall}) comprises two main components: f-InfoED for audio deepfake detection, and AdaLAM for feature extraction.
We also built ADFF 2024 dataset to further show the model's generalizability.

\vspace{-0.2cm}
\subsection{f-InfoED}
\label{sec:infoed}

Since directly applying entropy for deepfake detection faces challenges due to sensitivity to noise and content variations (depicted in Fig. \ref{fig:entropy_distribution}), we propose frame-level latent information entropy detector(f-InfoED).
This method leverages the variational information bottleneck \cite{alemi2016deep} to compress audio into a latent distribution, which contains the minimum information necessary for audio reconstruction.
This latent distribution, acts as a bottleneck, potentially mitigating noise effects by allowing only relevant information to pass through.
Also, we calculate frame-level latent entropy to capture information at fine-grained temporal intervals.
Our hypothesis is that inter-frame entropy differences may reflect information amount variations, thereby enabling the detection of deepfake artifacts.

Based on these considerations, we formulate two optimization objectives:
(1) Maximize the mutual information between the reconstructed audio $\hat{X}$ and extracted representation $Z$, while constraining $Z$'s mutual information with the input.
This ensures the latent representation encapsulates essential information of the input audio while minimizing redundancy.
(2) Maximizes classification accuracy according to latent information entropy.
Therefore, the optimization objective can be defined as follows:
\noindent
\begin{equation}
	\begin{aligned}
		\max_{\phi, \theta, \kappa} \mathbb{E}_{p_{\phi}\left(Z\right)}\left[p_{\kappa}\left(Y|\mathrm{H}\left(Z\right)\right)\right] + \mathrm{I}\left(Z,\hat{X}; \theta\right) \quad \\ s.t. \quad \mathrm{I}\left(Z,X; \phi\right) \leq \mathrm{I}_{c},
	\end{aligned}
\end{equation}
where $X$ denote the input, $\hat{X}$ the reconstructed input, and $Y$ the prediction target.
We define the $\phi$, $\theta$, and $\kappa$ as the parameters of neural networks representing $p_{\phi}$,$p_{\kappa}$, $p_{\theta}$, respectively.
The latent representation $Z$ is defined as $Z = \left\{z_i\right\}_i^C$, where $z_i \in \mathbb{R}^K$.
In this formulation, $C$ represents the frame sequence length, and $K$ denotes the latent representation dimensionality.

Employing Lagrange multipliers, we reformulate the objective function as:
\begin{equation}
	\max_{\phi, \theta, \kappa} \mathbb{E}_{p_{\phi}\left(Z\right)}\left[p_{\kappa}\left(Y|\mathrm{H}\left(Z\right)\right)\right] + \mathrm{I}\left(Z,\hat{X}; \theta\right)  - \beta \cdot \mathrm{I}\left(Z,X;\phi\right).
\end{equation}

Following \cite{higgins2017beta}, we exploited a probability distribution $q_{\phi}\left(Z|X\right)$ to approximate the posterior of latent embedding $Z$ in $p\left(Z|X\right)$, and we assumed that $p\left(Z|X\right)$ follows the Gaussian distribution.
Consequently, we express the first term can as
$\mathbb{E}_{q_{\phi}\left(Z|X\right)}\left[\log p_{\kappa}\left(Y|\mathrm{H}\left(Z\right)\right)\right]$.
As demonstrated by \cite{alemi2016deep}, we rewrite the second term as:
$\mathbb{E}_{q_{\phi}\left(Z|X\right)}\left[\log p_{\theta}\left(\hat{X}|Z\right)\right]$,
and bound the third term by $\mathrm{I}\left(Z,X\right) \leq  \mathrm{KL}\left[q_{\phi}\left(Z|X\right),r\left(Z\right)\right]$.
This yields the following lower bound:
\begin{equation}
	\begin{aligned}
		L = & \mathbb{E}_{q_{\phi}\left(Z|X\right)}  \left[\log q_{\kappa}\left(Y|\mathrm{H}\left(Z\right)\right)\right] + \mathbb{E}_{q_{\phi}\left(Z|X\right)}\left[ \log p_{\theta}\left(\hat{X}|Z\right)\right] \\
		    & - \beta \cdot \mathrm{KL}\left[q_{\phi}\left(Z|X\right),r\left(Z\right)\right],
	\end{aligned}
	\label{eq:vbi}
\end{equation}
where $r\left(Z\right)$ is assumed to follow a Gaussian distribution.
The formulation resembles the $\beta$-VAE and can be optimized using the re-parameterization trick.
Assuming the distribution $q_{\phi}\left(Z|X\right) \sim \mathcal{N}\left(f_{\mu}\left(x\right), f_{\Sigma}\left(x\right)\right)$ follows a Gaussian distribution, we can express the entropy calculation equation as:
\begin{equation}
	\begin{aligned}
		\mathrm{H}\left(Z\right) = \mathrm{H}\left(q_{\phi}\left(Z|X\right)\right) & = -\int q_{\phi}\left(Z|X\right) \ln \left(q_{\phi}\left(Z|X\right)\right) \mathrm{d}z \\
		                                                                           & =\frac{k}{2} \left(\ln2 \pi + 1\right) + \frac{1}{2}\ln |\Sigma|,
	\end{aligned}
\end{equation}
where $k$ denotes the dimension of the normal distribution, and $\Sigma$ is a diagonal matrix, meaning covariance.
The determinant in the second term poses a significant challenge for calculation and gradient computation.
To simplify, we assume the latent variable $Z$ follows a Gaussian distribution with a diagonal covariance matrix, such that $q_{\phi}\left(Z|X\right) \sim N \left(\mathbf{\mu}, \mathbf{\sigma}^{2}I\right)$, yielding:
\begin{equation}
	\begin{aligned}
		\ln |\Sigma| & = \ln \prod_{i=1}^{k} \sigma_{i}^{2}
		= 2 \cdot \sum_{i=1}^{k} \ln \sigma_{i}.
	\end{aligned}
\end{equation}
Here, $\sigma_{i}$ denotes the variance along the $i^{th}$ dimension.
Disregarding the constant term, we can calculate the entropy of $q_{\phi}\left(Z|X\right)$ as:
\begin{equation}
	\begin{aligned}
		\mathrm{H}\left(q_{\phi}\left(Z|X\right)\right)  \approx \sum_{i=1}^{k} \ln \sigma_{i},
	\end{aligned}
	\label{eq:entropy}
\end{equation}
This approximation enables efficient computation and gradient calculation.
Finally, we derive the frame-level entropy of the latent representation, which quantifies the information content of the compressed latent embedding.

As depicted in Fig. \ref{fig:overall}, our audio deepfake detection pipeline utilizes features extracted by AdaLAM as input $X$, aiming to reconstruct the audio $\hat{X}$ and predict the label $Y$ (bonafide/spoof).
These features are processed through mean encoder $f_{\phi}^{\mu}$ and variance encoder $f_{\phi}^{\Sigma}$ to obtain $\mathbf{\mu}$ and $\mathbf{\sigma}$, respectively.
Each encoder consists of a one-dimension convolution layer, an activation layer, and another one-dimension convolution layer, with $\mathbf{\mu}$ and $\mathbf{\sigma}$ both set to 192 dimensions.
We compute entropy using Eq. \ref{eq:entropy}, and derive the latent embedding $Z$ as follows:
\noindent
\begin{equation}
	z =  f^{\mu}_{\phi}\left(x\right)+f^{\Sigma}_{\phi}\left(x\right)\cdot\epsilon,
	\label{eq:z}
\end{equation}
\noindent
where $\epsilon$ was sampled from the normal distribution.
Subsequently, the latent embedding $Z$ is fed into a decoder, denoted as $p_{\theta}(\hat{X}|Z)$, which adopts an architecture similar to HiFi-GAN \cite{kong2020hifi}.
We configure the upsampling rates as 5, 4, 4, 2, 2, respectively, while maintaining other hyperparameters consistent with the original implementation.
The classifier, represented as $p_{\kappa}\left(\mathrm{H}(Z)\right)$, comprises two fully connected (FC) layers interspersed with an activation layer.
\vspace{-0.1cm}
\subsection{Loss Functions}
\label{sec:loss}

The entire network is trained using a multi-task learning approach, aiming to maximize the objective function defined in Eq. \ref{eq:vbi}.
The first term of Eq. \ref{eq:vbi} can be interpreted as the cross-entropy between the predicted target and the ground truth, denoted as $\mathcal{L}_{\mathrm{cls}}$, while the second term can be replaced by the reconstruction loss $\mathcal{L}_{\mathrm{recon}}$.
The training process involves the optimization of three loss functions: reconstruction loss $\mathcal{L}_{\mathrm{recon}}$, KL loss $\mathcal{L}_{\mathrm{KL}}$ in Eq. \ref{eq:lkl}, and classification loss $\mathcal{L}_{\mathrm{cls}}$. We exploited weighted cross entropy as the $\mathcal{L}_{\mathrm{cls}}$, and the weight was empirically set to 0.9 based on the label distribution of dataset.
For $\mathcal{L}_{\mathrm{recon}}$, we utilize the mean squared error (MSE) between the mel-spectrogram of the reconstructed audio $\mathbf{mel}_{\mathrm{pred}}$ and the ground truth $\mathbf{mel}_{\mathrm{gt}}$.
The Kullback-Leibler (KL) divergence between $q_{\theta}\left(Z|X\right)$ and normal distribution can be calculated as:
\begin{equation}
	\mathcal{L}_{\mathrm{KL}}=-\frac{1}{2}\left(\log \mathbf{\sigma}^{2} + 1 - \mathbf{\mu}^{2} - \mathbf{\sigma}^{2}\right).
	\label{eq:lkl}
\end{equation}
The total loss function $\mathcal{L}$ is the summation of  $\mathcal{L}_{\mathrm{recon}}$, $\mathcal{L}_{\mathrm{KL}}$ and $\mathcal{L}_{\mathrm{cls}}$ with the corresponding coefficent $\alpha$, $\beta$, and $\gamma$ with:
\begin{equation}
	\mathcal{L}=\alpha \cdot \mathcal{L}_{\mathrm{recon}}+\beta \cdot \mathcal{L}_{\mathrm{KL}}+\gamma \cdot \mathcal{L}_{\mathrm{cls}}.
\end{equation}

\subsection{AdaLAM: Large Audio Model with Trainable Adapters}
\label{sec:adawavlm}

To maintain consistency with an established audio deepfake detection pipeline, we adopted a two-component approach comprising a feature extractor and a classifier.
The performance of our f-InfoED is heavily influenced by the quality of the input features, $X$, which should encapsulate rich and essential audio information.

Recent studies have shown the efficacy of leveraging pre-trained large audio models \cite{chen2022wavlm, baevski2020wav2vec} for feature extraction \cite{guo2024audio, lv2022fake}, achieving better performance.
However, not all pre-trained models are inherently suitable for audio deepfake detection tasks.
Although fine-tuning has been explored as a potential solution, it presents challenges such as large amount of GPU memory requirements and the risk of catastrophic forgetting, potentially reducing generalization capability.

To address these challenges, we introduce an adapter layer to a pre-trained large audio model.
This adapter facilitates fine-tuning of fewer parameters and mitigates the catastrophic forgetting risk.
The adapter is inserted before the normalization layer in each of the transformer encoders (As shown in Fig. \ref{fig:overall}).
The adapter architecture comprises a residual bottleneck structure with two FC layers.
The first layer reduces the dimensionality, while the second restores it, with the residual output subsequently added back to the input.
In our implementation, we set the adapter projection dimension to 256.



\begin{table}[t]
	\setlength{\tabcolsep}{9.7pt}
	\renewcommand{\arraystretch}{1.0}
	\caption{Comparative EER (\%) results of AdaLAM \& f-InfoED with other methods in the ASVspoof 2021 DF dataset(DF) and In-the-wild. * denotes the model was trained by us.}
	\begin{tabular}{llcc}
		\toprule
		\textbf{Methods}                                       & \textbf{DF}   & \textbf{In-the-wild} \\
		\midrule
		CQTSpec \& Transformer \cite{muller2022does}           & /             & 43.78                \\
		LOGSpec \& MesoInception \cite{muller2022does}         & /             & 37.41                \\
		Wav2vec2-XLS-R \& AFN \cite{yi2023audio}               & 14.15         & 42.46                \\
		Wav2vec2XLS-R \& GAT \cite{yi2023audio}                & 14.91         & 44.31                \\

		OGSpec \& ASSERT \cite{yi2023audio}                    & 24.67         & 35.32                \\
		Wav2vec2-XLS-S \& ASSERT \cite{yi2023audio}            & 21.58         & 41.10                \\
		RawNet2 (End to End) \cite{tak2021end}                 & 24.32         & 42.14                \\

		AASIST (End to End) \cite{jung2022aasist}              & 18.75         & 40.46                \\
		XLS-R,WavLM,Hubert \& Fusion \cite{yang2024robust}     & /             & 24.27                \\
		XLS-R(ft) \& OCKD \cite{lu2024one}                     & 12.21*        & 33.17*               \\
		WavLM(ft) \& MultiFusion Attention \cite{guo2024audio} & 10.38*        & 31.58*               \\

		\midrule
		AdaLAM \& f-InfoED \textbf{(Ours)}                     & \textbf{9.24} & \textbf{8.36}        \\
		\bottomrule
		\label{tab1}
	\end{tabular}
	\vspace{-2em}
\end{table}

\section{Experiments}
\subsection{Experiments settings}
\subsubsection{Datasets}
To validate the generalizability of AdaLAM and f-InfoED, we conducted a comprehensive evaluation.
The model was initially trained on the publicly available ASVspoof 2019 LA training set, comprising 25,380 utterances.
Subsequently, we evaluated its performance across several test sets: the ASVspoof 2021 DF dataset, the In-the-wild dataset, and our newly proposed ADFF 2024 dataset. ADFF 2024 is a comprehensive audio deepfake dataset with 9 subsets, featuring various TTS and VC methods, totaling 22.48 hours of audio, designed to improve generation method descriptions and evaluation standards.
(The detailed description of the ADFF 2024 dataset is provided in Appendix A.).

\subsubsection{Metrics}
We mainly use equal error rate (EER) as the evaluation metric, which was selected due to its reliability in reflecting the standalone performance of the detection model.

\subsubsection{Implementation details}
Audio recordings were standardized to 4-second clips (64,600 samples), with longer clips truncated and shorter ones filled by repetitive padding.
To enhance the robustness, we employed RawBoost \cite{tak2022rawboost} for data augmentation.
All experiments were conducted on a single NVIDIA 4060ti GPU.
Detailed experimental parameters are provided in Appendix B.




\begin{table}[t]
	\setlength{\tabcolsep}{10.9pt}
	\renewcommand{\arraystretch}{1.0}
	\caption{Comparative EER (\%) results of our proposed method against other anti-spoofing systems based on the ADFF 2024 dataset, which comprises eight subsets, each representing deepfake audio generated by various TTS/VC methods.}
	\footnotesize
	\label{tab:af2024}
	\begin{tabular}{lcccccccc}
		\toprule
		\multicolumn{1}{l}{\multirow{1}{*}{\textbf{Subsets}}} & RawNet2 & AASIST & MFA   & \textbf{Ours}  \\
		\midrule
		FastSpeech2                                           & 2.32    & 0.43   & 13.26 & \textbf{0.40}  \\
		Bert-VITS                                             & 15.13   & 10.10  & 28.40 & \textbf{0.03}  \\
		GradTTS                                               & 12.52   & 2.07   & 38.4  & \textbf{0.20}  \\
		Bark                                                  & 15.63   & 13.20  & 14.23 & \textbf{1.47}  \\
		VALL-E (TTS)                                          & 39.44   & 43.33  & 22.63 & \textbf{5.40}  \\
		GPT-Sovits                                            & 36.25   & 17.23  & 34.13 & \textbf{5.67}  \\
		VALL-E (VC)                                           & 38.46   & 34.80  & 25.53 & \textbf{10.23} \\
		So-VITS-SVC                                           & 38.69   & 32.56  & 29.21 & \textbf{6.89}  \\
		\midrule
		Avg.                                                  & 24.81   & 19.22  & 25.72 & \textbf{3.79}  \\
		\bottomrule
	\end{tabular}
\end{table}

\subsection{Generalizability to Public Unseen Datasets}

To validate the generalizability of the proposed method, we conducted comparative experiments against current SOTA approaches under the same experimental setting.
For AASIST \cite{jung2022aasist}, MultiFusion Attention (MFA) \cite{guo2024audio}, and RawNet2 \cite{tak2021end}, which lacked reported performance on the In-the-wild dataset, we retrained these models using the same configuration as described in ASVspoof 2019 LA.
For other comparison methods, we directly cited performance metrics from their respective publications.
As shown in Table \ref{tab1}, our method demonstrated significant improvement on the In-the-wild dataset, achieving an EER of 8.36\%, which is significantly outperforms the second-best method at 65.6\%.
Furthermore, our approach achieved better result on the ASVspoof 2021 DF dataset, demonstrating its excellent generalization capabilities.

\subsection{Generalizability to the Latest Audio Generation Methods (ADFF 2024)}
To further evaluate the generalization capability of the proposed method, we evaluated our method on the ADFF 2024 dataset, which contains the latest VC and TTS methods.
Our model and baseline approaches were trained on the ASVspoof 2019 LA dataset and then evaluated on the various generation methods within ADFF 2024.
Table \ref{tab:af2024} shows that our method always achieved lower EER compared to current SOTA methods.
Notably, our method obtained EERs below 10\% for most subsets, except for a particularly challenging subset generated by VALL-E (VC).
Nevertheless, our approach still outperformed all other methods on this subset.
This performance analysis offers an additional perspective: our approach potentially provides a quantitative measure for assessing the efficacy of various audio generation methods.

\subsection{Generalizability to Unseen Perturbations}

Robustness to unexpected perturbations is a crucial issue, especially in the real world where audio data is often subject to various forms of interference or corruption.
In this study, we specifically chose two corruption scenarios to evaluate the robustness of our method: various audio durations and MP3 compression at different bit rates.
For the duration perturbations, we randomly segmented the raw audio into three different lengths (2 seconds, 3 seconds, and 4 seconds) and then padded them to a fixed 4-seconds length.
Furthermore, we evaluated the performance of our method under three MP3 compression bit rates: 115 kbps, 165 kbps, and 190 kbps.
We investigated the robustness of the proposed method and without f-InfoED.
As shown in Appendix C (Fig. 1), the performance degrades as the duration and bit rate of the input audio decrease.
However, our method consistently outperforms the current advanced approaches, even under the challenging conditions, such as audio durations as short as 2 seconds or very low bit rates at 115 kbps.

\subsection{Generalizability to Other Modality}
To demonstrate the broad applicability of InfoED in improving generalization across modalities, we extended our evaluation to the domain of image deepfake detection.
For this task, we employed ResNet-50 as the backbone architecture.
Since images lack the temporal dimension $t$ (in Eq. \ref{eq:entropy}) present in time-series data, we replace the calculation of latent information entropy by a learnable linear function of variance, $\sigma$. All other components keep consistent with f-InfoED.
For comparative analysis, we utilized images generated exclusively by ADM for training, while testing was conducted on images generated by various diffusion models, including ADM \cite{dhariwal2021diffusion}, DDPM \cite{ho2020denoising}, iDDPM \cite{nichol2021improved}, PNDM \cite{liu2022pseudo} and SD-V1 \cite{rombach2022high}, respectively.
Both training and testing datasets were sourced from \cite{wang2023dire}, and the training detail follows \cite{wang2023dire}.
The results (shown in Table \ref{tab:on_images}) provide compelling evidence of InfoED's potential to enhance generalizability for different modalities.

\begin{table}[t]
	\setlength{\tabcolsep}{6.2pt}
	\renewcommand{\arraystretch}{1.0}
	\caption{The accuracy (\%) of InfoED on image deepfake detection. The training images are generarated by ADM and test images are generated by other diffusion models \cite{wang2023dire}.}
	\begin{tabular}{cccccc}
		\toprule
		\textbf{Method}
		                  & ADM          & DDPM          & iDDPM        & PNDM          & SD-V1         \\
		\midrule
		RestNet-50        & 100          & 87.3          & 100          & 77.8          & 77.4          \\
		RestNet-50+InfoED & \textbf{100} & \textbf{93.4} & \textbf{100} & \textbf{85.2} & \textbf{86.3} \\
		\bottomrule
		\label{tab:on_images}
	\end{tabular}
	\vspace{-2.6em}
\end{table}

\subsection{Analytical Studies of Our Proposed Method}
\subsubsection{Effect of f-InfoED and AdaLAM}
To evaluate the generalized efficacy of each module in the proposed approach, we conducted comprehensive ablation studies using the same benchmark setup, training on the ASVspoof 2019 LA dataset, and testing on the In-the-wild dataset. Three ablated configurations were employed as contrasting baselines:
(1) The first configuration removed the reconstruction loss $\mathcal{L}_{\mathrm{recon}}$, thereby eliminating constraint ensuring the latent embedding $Z$ can accurately respresents the audio.
(2) The second configuration removed both the reconstruction loss $\mathcal{L}_{\mathrm{recon}}$ and KL loss $\mathcal{L}_{\mathrm{KL}}$, resulting in a traditional classifier architecture consisting of two 1-D convolutional layers and two fully connected layers.
(3) The third configuration replaced the AdaLAM by the pretrained WavLM base+ model.
Table \ref{tab:ablation_infoed} presents the comparative results of several combination of these configuration, providing strong evidence for the effectiveness of the f-InfoED and AdaLAM approach in generalized audio deepfake detection.



\subsubsection{Selection of the Hyper-parameters:}
The hyperparameter selection process focused on three key parameters: the weight coefficients among losses (in Eq. \ref{eq:lkl}) with $\alpha$, $\beta$, and $\gamma$.
We implemented a two-stage hyperparameter tuning procedure.
Initially, we fixed $\gamma=1$ and varied $\alpha$ and  $\beta$ across multiple trials.
Subsequently, we fixed the optimized values of $\alpha$ and $\beta$ and explored variations in $\gamma$.
As shown in Appendix C (Table 2), our method exhibited optimal performance with the hyperparameter configuration of $\alpha=0.95$, $\beta=0.05$, and $\gamma=1$.
These results show the importance of each loss and demonstrate the robustness of our method to hyperparameter variations, maintaining stable performance under different configurations.

\subsubsection{Visualisation of Entropy and Extracted features:}
\begin{figure}[t]
	\centering
	\includegraphics[width=0.93\linewidth]{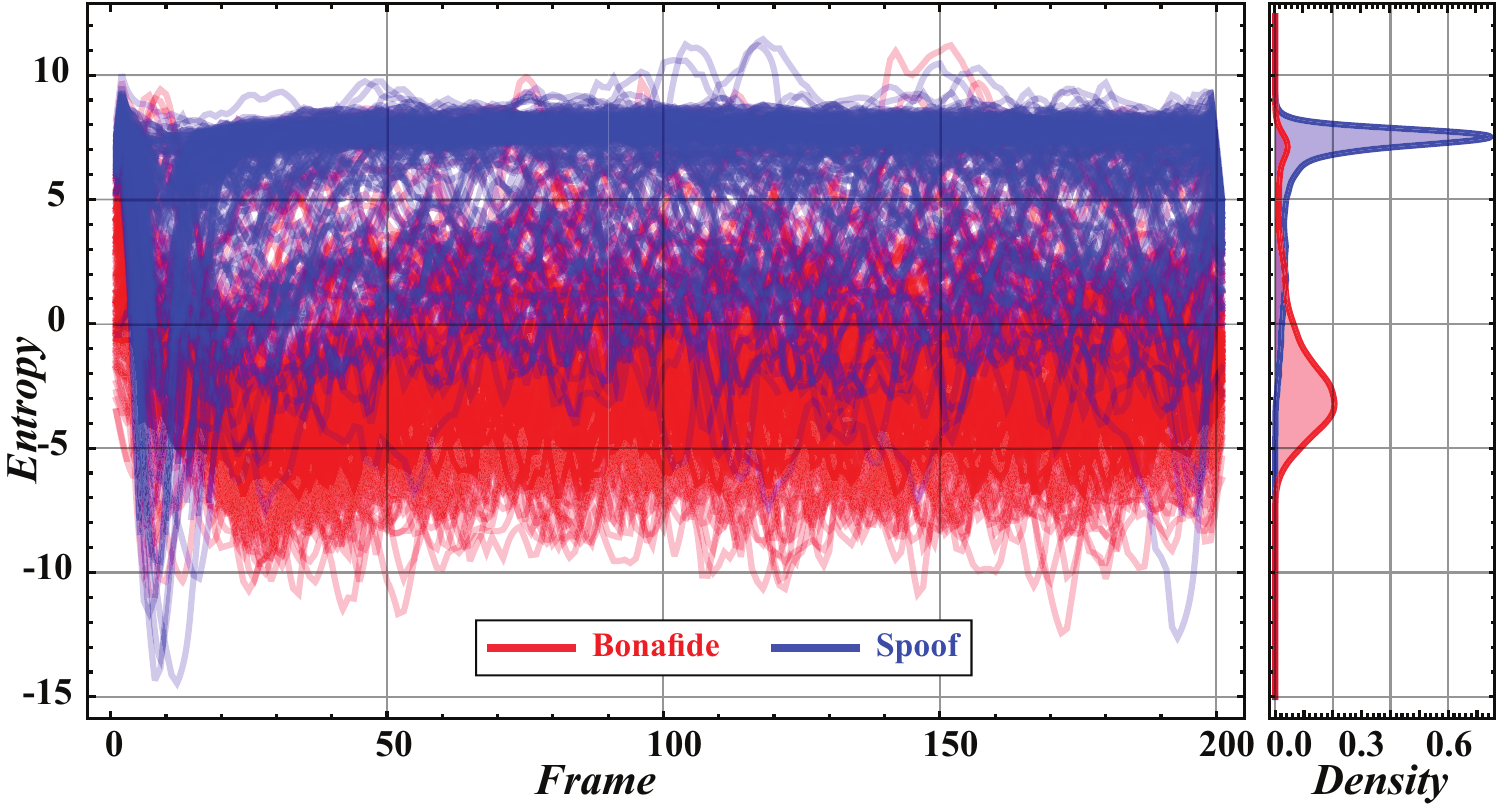}
	\vspace{-0.25cm}
	\caption{The visualization of the frame-level information entropy for the In-the-wild dataset.}
	\label{fig:entropy}
\end{figure}

\begin{figure}[t]
	\centering
	\includegraphics[width=0.93\linewidth]{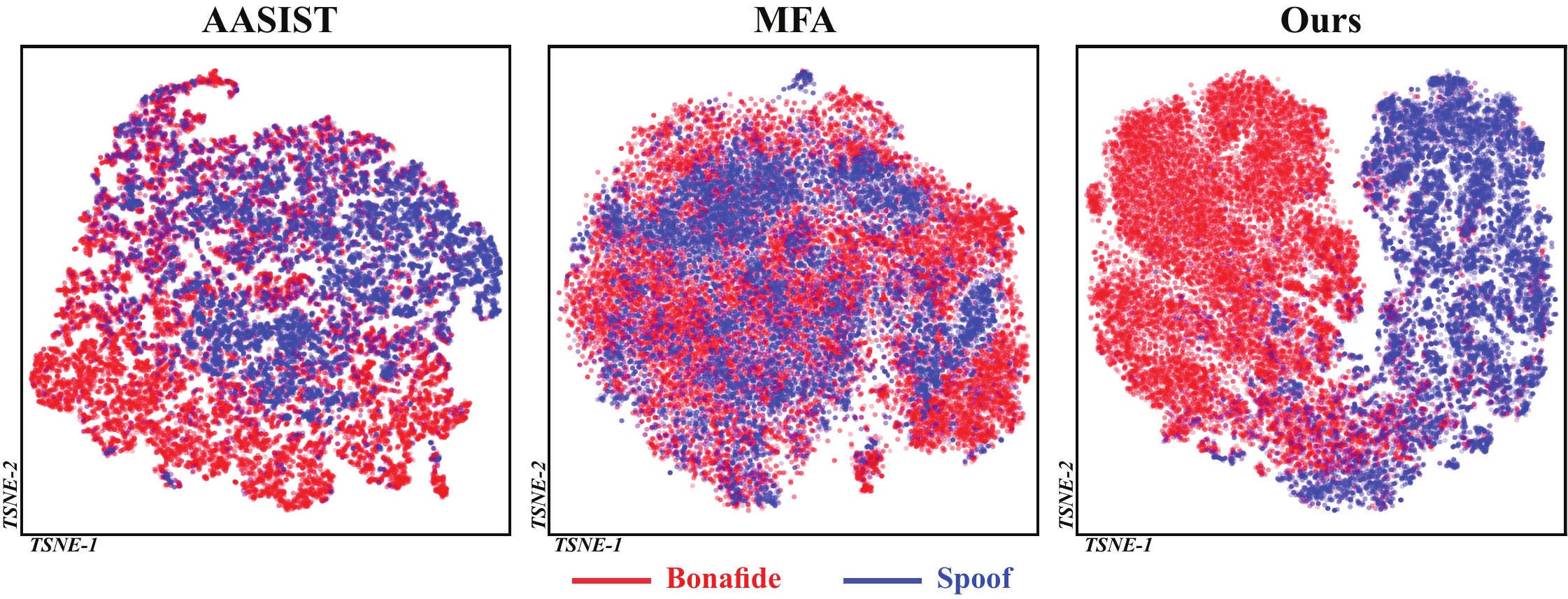}
	\vspace{-0.5em}
	\caption{The t-SNE plot of AASIST, MFA, and our proposed method on the latent space for the In-the-wild dataset.}
	\label{fig:tsne}
\end{figure}

To elucidate the intrinsic mechanism of our proposed method, we conducted visualization experiments on the In-the-wild dataset.
We computed and visualized the latent information entropy, as illustrated in Fig. \ref{fig:entropy}.
The analysis revealed that spoofed audio samples exhibited higher entropy compared to bonafide audio, suggesting that the bonafide audio contains more informative content.
This observation aligns with our findings on the ASVSpoof 2019 / 2021 datasets (Fig. \ref{fig:entropy_distribution}) and is consistent with the motivation for our approach.
Interestingly, we observed the entropy mixed between bonafide and spoofed samples mixed at the initial part of the audio samples, potentially due to short silence intervals at the beginning of the audio.
Furthermore, we employed t-SNE to visualize the deep feature representations extracted by our method and compared them with the AASIST and MFA baselines (Shown in Fig. \ref{fig:tsne}).
The results demonstrate that our approach generates feature representations with superior discriminative properties for bonafide and spoofed audio categories.
This suggests that our method extracts more informative and discriminative features, leading to enhanced classification performance.

\begin{table}[t]
	\centering
	\caption{Ablation Experimental Results on In-the-Wild.}
	\footnotesize
	\setlength{\tabcolsep}{17.5pt}
	\renewcommand{\arraystretch}{1}
	\begin{tabular}{ccccc}
		\toprule
		$\mathcal{L}_{\text{recon}}$ & $\mathcal{L}_{\mathrm{KL}}$ & \textbf{AdaLAM} & \textbf{EER (\%)} $\downarrow$ \\
		\midrule
		                             &                             &                 & 32.13                          \\
		$\checkmark$                 & $\checkmark$                &                 & 21.53                          \\
		                             & $\checkmark$                & $\checkmark$    & 14.19                          \\
		                             &                             & $\checkmark$    & 12.76                          \\
		$\checkmark$                 & $\checkmark$                & $\checkmark$    & 8.360                          \\
		\bottomrule
	\end{tabular}
	\label{tab:ablation_infoed}
\end{table}

\noindent

\section{Conclusion}

We introduced a novel method in this paper for generalized audio deepfake detection with two key components:
(1) AdaLAM, which extends pre-trained audio models with trainable adapters for more discriminative feature extraction.
(2) The f-InfoED, which compresses these features into frame-level latent information entropy, serves as a robust indicator.
This information-theoretic perspective allows our method to generalize well to unseen deepfake samples, overcoming the limitations of existing detection approaches.
To facilitate comprehensive evaluation, we have built the ADFF 2024 dataset generated by the latest TTS and VC methods.
Experiments demonstrate that our proposed framework achieves SOTA performance with remarkable generalization capabilities and is efficient for unseen perturbation and other modalities.
We believe that this work will stimulate further research from an information-theoretic perspective, not only for audio deepfake detection but also for other research areas and applications.
\vspace{0.13cm}


\section*{Acknowledgment}
This work was supported by the National Key Research and Development Program of China (Youth Scientist Project, Grant No. 2024YFB4504300), the Key Research and Development Program of Guangdong Province (grant No. 2021B0101400003) and the corresponding author is Jianzong Wang (jzwang@188.com).

\appendix

\subsection{The Detail of ADFF 2024}
The \textbf{A}udio \textbf{D}eep\textbf{f}ake \textbf{F}orensics \textbf{2024} (ADFF 2024) consists of 9 subsets, and for two major tasks: text-to-speech (TTS) and voice conversion (VC).
\subsubsection{Text-to-Speech}
For this study, we compiled a dataset of synthesized audio samples from seven state-of-the-art TTS models: FastSpeech2\footnote{\Href{https://github.com/ming024/FastSpeech2} \label{FastSpeech2}},
Bert-VITS2\footnote{\Href{https://github.com/fishaudio/Bert-VITS2} \label{Bert-VITS2}},
VALL-E\footnote{\Href{https://github.com/Plachtaa/VALL-E-X} \label{VALL-E}},
Bark\footnote{\Href{https://github.com/suno-ai/bark} \label{Bark}},
GradTTS\footnote{\Href{https://github.com/huawei-noah/Speech-Backbones/tree/main/Grad-TTS} \label{GradTTS}},
and GPT-SoVITS\footnote{\Href{https://github.com/RVC-Boss/GPT-SoVITS} \label{GPT-SoVITS}}.
These models represent a diverse range of TTS approaches, from non-autoregressive methods to recent language modeling-based techniques.

FastSpeech2 \cite{ren2020fastspeech} is a well-known non-autoregressive TTS method that incorporates predictors for duration, energy, and pitch, enabling the generation of diverse speech characteristics.
GradTTS \cite{popov2021grad} introduces a novel approach utilizing a score-based decoder to produce high-fidelity speech.
Bert-VITS (BVITS) \cite{kim2021conditional}, an extension version of the VITS model, has gained significant attention in the open-source community due to its performance improvements.
VALL-E \cite{wang2023neural}, Bark, and GPT-SoVITS represent the cutting edge in TTS technology, employing language modeling approaches that demonstrate emerging in-context learning capabilities.
These models excel in synthesizing high-quality personalized speech with minimal speaker-specific data as an acoustic prompt.

To generate the synthetic audio samples for our analysis, we selected text passages at random from the multi-speaker VCTK dataset.
The pre-trained models were employed using their default configurations, with speaker identities randomly assigned for each generated sample.

\begin{table}[t]
	\centering
	\setlength{\tabcolsep}{11.65pt}
	\renewcommand{\arraystretch}{1.0}
	\caption{The composition of the ADFF 2024 dataset.}
	\begin{tabular}{@{}lccc@{}}
		\toprule
		\textbf{Generator} & {\textbf{Total Hours}} & \textbf{Average Duration (s)} & \textbf{Type}        \\ \midrule
		Ground Truth       & 2.80                   & 3.36                          & /                    \\ \midrule
		FastSpeech2        & 2.60                   & 3.12                          & \multirow{6}{*}{TTS} \\
		Bert-VITS          & 1.87                   & 2.24                          &                      \\
		VALL-E (TTS)       & 2.55                   & 3.06                          &                      \\
		Bark               & 3.56                   & 4.27                          &                      \\
		GradTTS            & 2.34                   & 2.81                          &                      \\
		GPT-SoVITS         & 2.36                   & 2.83                          &                      \\ \midrule
		VALL-E (VC)        & 2.22                   & 2.66                          & \multirow{2}{*}{VC}  \\
		So-VITS-RVC        & 2.18                   & 2.82                          &

		\\ \midrule
		Overall            & 22.48                  & 3.00                          & /
		\\ \bottomrule
	\end{tabular}
	\label{tab:dataset}
\end{table}

\subsubsection{Voice Conversion}
Additionally, we expanded our dataset to include audio generated through VC methods.
Specifically, we utilized VALL-E\textsuperscript{\ref{VALL-E}} and So-VITS-SVC\footnote{\Href{https://github.com/svc-develop-team/so-vits-svc} \label{sovits}} for this purpose.
VALL-E, while primarily known for its TTS capabilities, also demonstrates proficiency in VC tasks.
So-VITS-SVC, a derivative of the VITS model, has gained considerable popularity within the open-source community for its effectiveness in singing voice conversion (SVC).

For VALL-E, our approach involved randomly selecting audio samples from two different speakers within the VCTK dataset.
One sample served as the reference voice, while the other provided the input content for conversion.
This method allows for diverse speech transformations.
For So-VITS-SVC, the pretrained model is selected, which is trained on the VCTK dataset, to generated spoofed audio samples by converting randomly selected samples from the dataset.

In total, the ADFF 2024 dataset comprises 22.48 hours of audio material, with individual samples averaging 3 seconds in duration.
A comprehensive breakdown of the dataset composition is provided in Table \ref{tab:dataset}.
This dataset represents current advanced synthetic speech samples covering a wide range of TTS and VC techniques, allowing for a more comprehensive assessment of the generality of the models for comparison purposes.

\subsection{Experimental Settings}
\subsubsection{Datasets}

To assess the generalization capabilities of our proposed method, we performed a comprehensive evaluation using multiple datasets.
The model was initially trained on the ASVspoof 2019 LA training dataset, a publicly available corpus containing 25,380 utterances.
Subsequently, we evaluated its performance on ASVspoof 2021 DF dataset, an In-the-wild dataset, and our generated ADFF 2024 dataset.
The ASVspoof 2019 LA dataset serves as a robust training set, containing both bonafide and spoofed utterances generated by 19 previously popular spoofing methods, including various TTS and VC techniques.
It was splitted into training, development, and evaluation sets, containing 6, 2, and 11 different spoofing methods, respectively.
For a more rigorous evaluation for model generalization, we employed the more challenging ASVspoof 2021 DF subset, which contains previously unseen audio coding and compression artifacts.
To further validate the model's performance under real-world conditions, we utilized an In-the-wild dataset, which is collected from various publicly accessible sources such as social networks and video-sharing platforms, consists of 19,963 bonafide and 11,816 spoofed utterances, offering a diverse range of naturally occurring audio samples.
In preprocessing the audio data, we split all audio in into 4-second clips of 64,600 samples, truncating longer recordings and filling shorter segments by repetitive padding.
To enhance the robustness and generalization capabilities of our model, we employed RawBoost \cite{tak2022rawboost} for data augmentation.

\subsubsection{Implementation details}

For calculating the reconstruction loss function, mel-spectrograms were derived from the audio.
Initially, a short-time Fourier transform (STFT) was applied with a hop size of 256 and a window size of 1024.
The resulting spectrum was then converted into an 80-dimensional mel-spectrogram.
This captured detailed frequency components for accurate reconstruction loss calculation.
We employed the WavLM base+ model as the pre-trained large audio model \cite{chen2022wavlm}, choicen for its proven effectiveness in various audio processing tasks.
The training process utilized the Adam optimizer with an initial learning rate of $3 \times 10^{-5}$ and a batch size of 8.
The best-performing model was selected based on ASVspoof 2019 development set after 50 training epochs.
All experiments were conducted on a single NVIDIA 4060ti GPU.

\subsubsection{Compared baselines}

To comprehensively evaluate our proposed method, we conducted comparisons with various state-of-the-art baselines.
Firstly, we assessed the performance using various feature extraction techniques, including Constant-Q Transform (CQTSpec) \cite{brown1991calculation}, log-spectrogram (LOGSpec), Wav2vec2 \cite{baevski2020wav2vec}, fine-tuned WavLM \cite{guo2024audio}, Wav2vec2 \cite{wang2023detection}, feature fusion approaches \cite{yang2024robust}, as well as our proposed AdaLAM.
Additionally, we compared our method with different classifiers, such as Transformer \cite{wang2023detection}, mel-spectrogram inception \cite{afchar2018mesonet}, attention filtering network (AFN) \cite{lai2019attentive}, graph attention network (GAT) \cite{tak2021graph}, anti-spoofing with squeeze-excitation and residual networks (ASSERT) \cite{lai2019assert}, and our proposed f-InfoED.
Furthermore, we evaluated end-to-end methods AASIST \cite{lai2019assert} and RawNet2 \cite{tak2021end} to provide a comprehensive comparison across different architectural approaches.

\subsection{Supplementary Results}
\begin{figure}[ht]
	\vspace{-1.5em}
	\centering
	\includegraphics[width=0.98\linewidth]{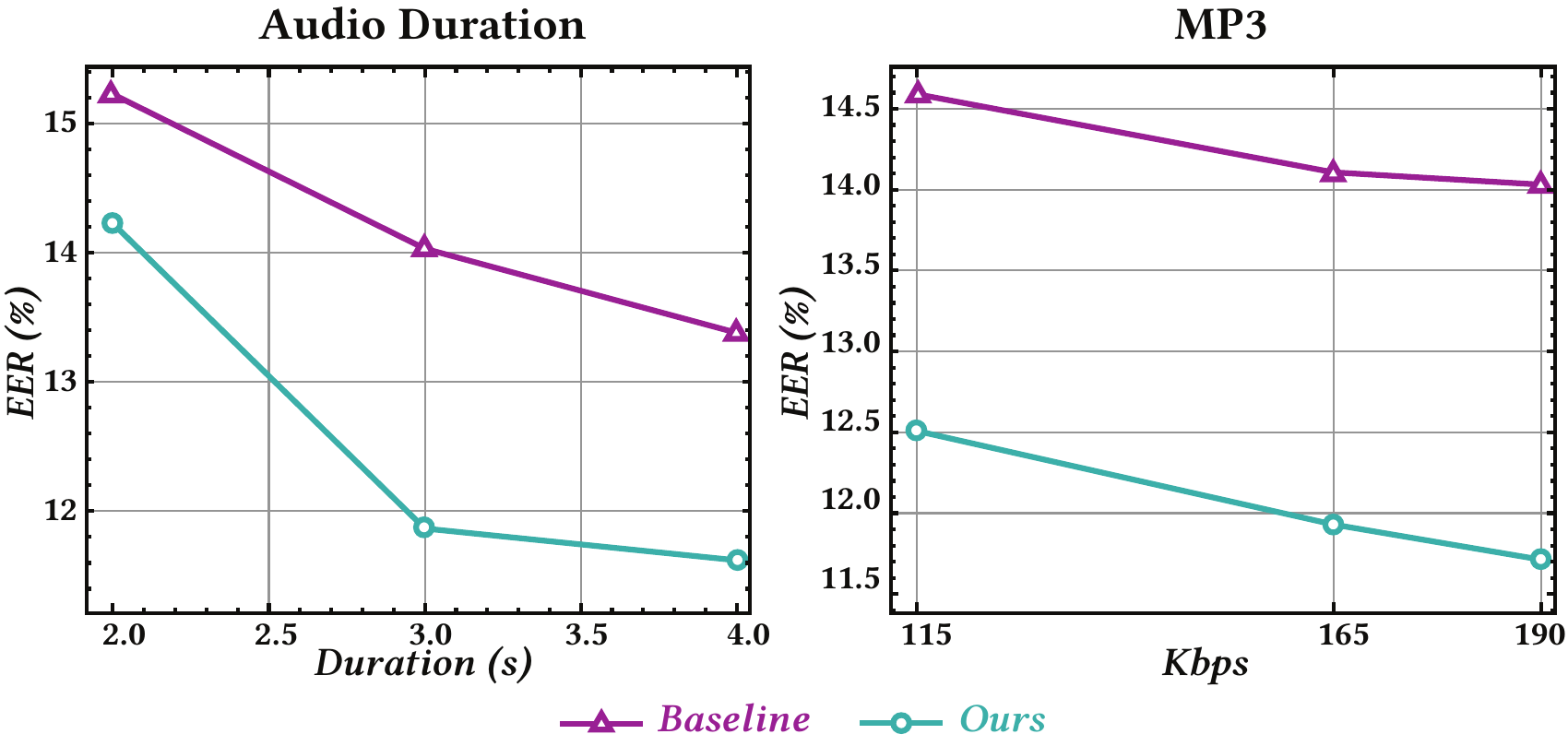}
	\caption{Robustness to unseen perturbations.
		It shows perturbations on the In-the-wild dataset for different audio durations and MP3 compression rates, and is reported as EER (\%) for robustness comparison.}
	\label{fig:perturbation}
\end{figure}

Due to space constraints in the main paper, there are few supplementary results shown in Table \ref{tab:hyper} and Figure \ref{fig:perturbation}.

\begin{table}[ht]
	\setlength{\tabcolsep}{29pt}
	\renewcommand{\arraystretch}{1.0}
	\caption{The ablation results of hyper-parameters on In-the-Wild. $\alpha$, $\beta$ and $\gamma$ are the coefficient coefficients of $\mathcal{L}_{\mathrm{recon}}$, $\mathcal{L}_{\mathrm{KL}}$ and $\mathcal{L}_{\mathrm{cls}}$, repectively.}
	\begin{tabular}{lc}
		\toprule
		\multicolumn{1}{l}{\multirow{1}{*}{\textbf{Configuration}}}
		                                          & \textbf{EER (\%)} \\

		\midrule

		$\alpha=0.80$, $\beta=0.20$, $\gamma=1$   & 10.08             \\
		$\alpha=0.90$, $\beta=0.10$, $\gamma=1$   & 9.24              \\
		$\alpha=0.95$, $\beta=0.05$, $\gamma=1$   & \textbf{8.36}     \\
		$\alpha=0.99$, $\beta=0.01$, $\gamma=1$   & 9.82              \\
		\midrule
		$\alpha=0.95$, $\beta=0.05$, $\gamma=1$   & \textbf{8.36}     \\
		$\alpha=0.95$, $\beta=0.05$, $\gamma=10$  & 9.22              \\
		$\alpha=0.95$, $\beta=0.05$, $\gamma=100$ & 10.31             \\
		\bottomrule
		\label{tab:hyper}
	\end{tabular}
\end{table}

\balance
\bibliographystyle{IEEEbib}
\bibliography{refs}

\end{document}